\documentclass[showpacs,twocolumn]{revtex4}%

\usepackage{graphicx}% Include figure files
\usepackage{dcolumn}% Align table columns on decimal point
\usepackage{bm, array}% bold math
\usepackage{multirow}%
\usepackage[dvipsnames]{xcolor}
\usepackage{longtable}
\usepackage{latexsym}

\def\sc{\texttt{\footnotesize sc}}

\def\mix{\texttt{\footnotesize mix}}
\def\gs{\texttt{\footnotesize gs}}

\def\at{\texttt{\footnotesize at}}
\def\Ti{\texttt{\footnotesize Ti}}
\def\ls{\texttt{\footnotesize ls}}
\def\nv{\texttt{\footnotesize nv}}
\def\eq{\texttt{\footnotesize eq}}
\def\dis{\texttt{\footnotesize dis}}

\def\mix{\texttt{\footnotesize mix}}
\def\sol{\texttt{\footnotesize sol}}
\def\A{\texttt{\footnotesize A}}
\def\B{\texttt{\footnotesize B}}
\def\D{\texttt{\footnotesize D}}
\def\V{\texttt{\footnotesize V}}
\def\sA{\texttt{\footnotesize A}}

\def\DEG{{$^\circ$C}}

\begin{document}
\textheight 226.5mm

\title{Calculation of solubility in titanium alloys from first-principles}

\author{Roman V. Chepulskii and Stefano Curtarolo\cite{SCemail}}

\affiliation{Department of Mechanical Engineering and Materials Science and Department of Physics,
  Duke University, Durham, NC 27708, USA}

\date{\today}

\begin{abstract}
  We present an approach to calculate the atomic bulk solubility in binary alloys
  based on the statistical-thermodynamic theory of dilute lattice gas.
  The model considers all the appropriate ground states of the alloy and
  results in a simple Arrhenius-type temperature dependence determined by a
  {\it ``low-solubility formation enthalpy''}.
  This quantity, directly obtainable from first-principle calculations,
  is defined as the composition derivative of the compound formation enthalpy with respect to
  nearby ground states.
  We apply the framework and calculate the solubility of the A specie in  A-Ti alloys
  (A=Ag,Au,Cd,Co,Cr,Ir,W,Zn).
  In addition to determining unknown low-temperature ground states for the eight alloys,
  we find qualitative agreements with solubility experimental results.
  The presented formalism, correct in the low-solubility limit, should be considered as an appropriate
  starting point for determining if more computationally expensive formalisms are otherwise needed.
\end{abstract}

%\pacs{}

\maketitle

\section{Introduction}

High-throughput {\it ab initio} methods, capable of predicting
properties of an ample set of materials from quantum mechanics
calculations, are becoming important tools for scientists working in
rational materials development. These methods allow researchers to
correlate between different systems and to observe trends converging
toward the predictions of new materials
\cite{Ceder1998Nature,Johannesson2002PRL,Crespi,Curtarolo2003PRL,Curtarolo2005Calphad,Ceder2006NatureMaterials}.
The main difference between the ``several-calculations'' and the
``high-throughput'' philosophies is that the latter requires rapid
estimations of materials properties so that the correlations between
systems, even if roughly characterized, become the target
information instead of the throughout and accurate description of a
small subset. Clearly, to analyze the extensive amount of
information and to extract correlations, ad-hoc algorithms and
appropriate computer softwares have to be developed. Furthermore,
once the space of the search is narrowed, a detailed study can be
employed on the obtained reduced set of feasible candidate systems.

Several examples have appeared in literature in recent years, for
instance the ``data-mining of quantum calculations'' method leading
to the principle component analysis of the formation energies of
many alloys in several configurations
\cite{Curtarolo2003PRL,Curtarolo2005Calphad}, the evolutionary
approach for determining hamiltonian \cite{Hart2005NatMat}, the
``Pareto-optimal'' alloys and catalysts
\cite{Bligaard2003APL,Andersson2006JCatal} the prediction of the
lithium-boron superconductor \cite{artSC21}, the ``high-throughput
Kohn-anomalies'' search in ternary lithium-borides
\cite{artSC33,artSC37}, and the ``multi-optimization'' techniques
used in studying high-temperature reactions in multicomponent
hydrides
\cite{Wolverton2008JPCM,Siegel2007PRB,Akbarzadeh2007AdvMat}.

This manuscript focuses on the high-throughput formalism for the
calculations of solubility in binary alloys (solvus lines). The
knowledge of solubility is crucial for designing new alloys with
particular physical, chemical, and mechanical properties. For
example, if we have to enhance an alloy property by adding extra
specie as solute, it is necessary to know the equilibrium solubility
to understand if it is possible to dissolve the candidate specie,
and, if possible, to avoid supersaturation-precipitation and
subsequent modification of the target property (aging effect). In
superconducting materials research, the problem emerges frequently:
often expensive and difficult experiments are undertaken to enhance
the critical temperature \cite{Cava2003phisicaC,Singh08RbCs}. Even
in catalysis research, recent experiments and modeling have shown
that the solubility of carbon is responsible of thermodynamic
instabilities hindering the catalytic activity of very small Fe and
Fe:Mo clusters \cite{artSC36,artSC39}. The calculation of solubility
of Zr in Al has already been addressed with success within the
regular solution model \cite{SigliPROC} fit to {\it ab initio}
calculations \cite{SanchezPRB}. Solubility can also be extracted
from the knowledge of the phase diagram which, in the case of
lattice-conserving alloys, can be generated within the the cluster
expansion \cite{CD_deFontaine,CE_Ceder} and Monte Carlo
approaches\cite{Anton1,Anton2}. However, a straightforward
formalism leading to the estimation of equilibrium solubility for
general alloys is still lacking.

In the present paper we devised a statistical-thermodynamic approach
for the calculation of atomic solubility in alloys. The advantage of
our approach consists in taking into account all available ground
states rather than just the pure species configurations. To test the
method, we present calculations for a number of binary titanium systems.
The paper is organized as follows: In Sec. II we rewrite the
equations governing solubility in the case of vacancies and
substitutional impurities in binary alloys. Section III is devoted
to the discussions of capabilities and limits of our formalism.
Examples of phase diagrams and solubilities are addressed in Sec. IV
for the following test Ti-A systems (A=Ag,Au,Cd,Co,Cr,Ir,W,Zn).
Conclusions are given in Sec. V.

\section{Solubility formalism}
\label{Solubility}
\section*{Enthalpy}

Let us consider a disordered dilute solution of A-atoms and
vacancies (V) in a pure B-solid with a given Bravais crystal lattice
(labeled with the {\it ``dis''}). Without taking into account the
A-A, A-V, and V-V interactions and assuming that A and V
concentrations are small, the approximate enthalpy of the considered
alloy can be written as:
\begin{equation}\label{Eq:H}
    H^\dis=E^\dis+pV^\dis=
    H_\B^\at N + H_{\A_\B} N_\A +
    H_{\V_\B} N_\V,
\end{equation}
where $H_\B^\at$, $H_{\A_\B}$ and $H_{\V_\B}$ are the enthalpy of
the pure B solid per unit cell, the change in enthalpy of the solid
upon substitution of one B with an A-atom, and the change
in enthalpy upon removal of one B atom, respectively. In
addition, $N$ and $N_\alpha$ ($\alpha$=A,B,V) are the total numbers
of crystal lattice sites and of atoms of $\alpha$-type:
\begin{equation} \label{Eq:H1}
    \begin{array}{c}
    N=N_\A+N_\B+N_\V,\,\,\,\,\,
    H_\B^\at=E_\B^\at+pv_\B^0,\\\\
    H_{\A_\B}=E_{\A_\B}^0+pv_{\A_\B}^0,\,\,\,
    H_{\V_\B}=E_{\V_\B}^0+pv_{\V_\B}^0,
    \end{array}
\end{equation}
$E_\alpha^\at$ and $v_\alpha^0$ ($\alpha$=A,B) are energy and the
volume per atom of the $\alpha$-pure solid, $p$ is the
pressure, $v_{\A_\B}$ and $v_{\V_\B}^0$ represent the change of
volume of the B-pure solid upon introduction of one A-atom
or one vacancy. The framework introduced by Eq.
(\ref{Eq:H}) is similar to Wagner-Schottky model of a system of
non-interacting particles.\cite{WagnerSchottky30}. The quantities
with superscript ``0'' are considered to be temperature, pressure,
and composition independent being calculated at zero temperature and
pressure. Hence, the accuracy of the following results will be
better in the case of limited pressures or for systems with very
high bulk modulus, where the elastic energy fraction of
$E_\alpha^\at$ is negligible.

The quantities in Eq. (\ref{Eq:H1}) can be easily approximated as
differences of first-principles energies and volumes between large
supercells (sc) with or without defects (substitutional A-atom or
vacancy):
\begin{equation} \label{Eq:H3}
    \begin{array}{c}
    E_{\A_\B}^0 \simeq E_\texttt{sc}[B_{n_\texttt{sc}-1}A]-E_\texttt{sc}[B_{n_\texttt{sc}}],\\
    v_{\A_\B}^0 \simeq V_\texttt{sc}[B_{n_\texttt{sc}-1}A]-V_\texttt{sc}[B_{n_\texttt{sc}}],\\\\
    E_{\V_\B}^0 \simeq E_\texttt{sc}[B_{n_\texttt{sc}-1}]-E_\texttt{sc}[B_{n_\texttt{sc}}],\\
    v_{\V_\B}^0 \simeq V_\texttt{sc}[B_{n_\texttt{sc}-1}]-V_\texttt{sc}[B_{n_\texttt{sc}}].
    \end{array}
\end{equation}
As the size of the supercell grows,
the approximate quantities in Eqs. (\ref{Eq:H3})
approach their exact values.
In literature\cite{Mishin} and in this paper,
$E_{\alpha_\B}^0$, $v_{\alpha_\B}^0$ and
$H_{\alpha_\B}$ ($\alpha$=A,V) are called the ``raw''
(composition unpreserving) $\alpha$-defect formation energy, volume
and enthalpy, respectively.

The enthalpy per atom is obtained from Eq. (\ref{Eq:H}) as:
\begin{equation} \label{Eq:H4}
    \begin{array}{c}
    H^\dis_\at=H^\dis/(N_\A+N_\B)=\\\\
    H_\B^\at + H_{\A_\B} x_\A +
    (H_{\V_\B}+H_\B^\at) x_\V,
    \end{array}
\end{equation}
where $x_\alpha$ ($\alpha$=A,B,V) are the atomic concentrations
\begin{equation}
x_\alpha=N_\alpha/(N_\A+N_\B).
\end{equation}
Then, we follow the convention of using the {\it formation}
enthalpy $\Delta H^\dis_\at$ calculated with respect to the pure A-
and B-solids \cite{Korzh00}:
\begin{equation} \label{Eq:H5}
    \Delta H^\dis_\at = H^\dis_\at-x_\A
    H_\A^\at-(1-x_\A)H_\B^\at,
\end{equation}
where $H_\A^\at=E_\A^\at+pv_\A^0$
and $H_\B^\at$ has been defined
in Eq. (\ref{Eq:H1}).
Combining equations (\ref{Eq:H4}) and (\ref{Eq:H5}), we get
\begin{equation}\label{Eq:H7}
    \Delta H^\dis_\at = H_\A x_\A+H_\V x_\V,
\end{equation}
where the quantities $H_\A$ and $H_\V$, defined as
\begin{equation}\label{Eq:H8}
    H_\A=H_{\A_\B}-H_\A^\at+H_\B^\at,
    \,\,\,\,\,\,\,\,H_\V=H_{\V_\B}+H_\B^\at
\end{equation}
are usually called A-defect and V-defect ``true'' (composition
preserving)\cite{Mishin} formation enthalpies, which can be
obtained from\cite{Korzh00}
\begin{equation}\label{Eq:H9}
    H_\alpha=\left. \frac{\partial \Delta H^\dis_\at}{\partial x_\alpha}\right | _{x_\alpha \rightarrow 0}
    \,\,\,\,  (\alpha={\rm A,V}).
\end{equation}

\section*{Equilibrium Gibbs Free Energy}

The formation Gibbs free energy, $\Delta G^\dis_\at$, is defined as:
\begin{equation}\label{Eq:G1}
    \Delta G^\dis_\at=\Delta
    H^\dis_\at-T \Delta S^\dis_\at,
\end{equation}
where $\Delta H^\dis_\at$ is described by Eq. (\ref{Eq:H7}) and the
formation entropy $\Delta S^\dis_\at=S^\dis_\at$ can be obtained
within the mean-field approximation as
\begin{equation}
\label{Eq:G2}
    \Delta S^\dis_\at= - \frac{k_\B N}{N_\A+N_\B} \sum_{\alpha=\texttt{A,B,V}}
    c_\alpha \ln c_\alpha,
\end{equation}
where $T$, $k_\B$, $c_\alpha$ ($\alpha$=A,B,V) are the temperature,
the Boltzmann constant, and the site concentrations of atoms:
\vspace{-4mm}
\begin{equation}\label{Eq:G3}
c_\alpha=N_\alpha/N.
\end{equation}
By changing variables from site concentrations $c_\alpha$ to atomic concentrations $x_\alpha$:
\begin{equation}\label{Eq:G4}
x_\alpha=N_\alpha/(N_\A+N_\B),\,\,\,\,\,\,c_\alpha=x_\alpha/(1+x_\V),
\end{equation}
we rewrite Eq. (\ref{Eq:G1}) as
\begin{equation}\label{Eq:G5}\label{Eq:G6}
\begin{array}{c}
    \Delta G^\dis_\at=\Delta G^{\dis,\A}_\at+\Delta G^{\dis,\V}_\at, \\\\
    \Delta G^{\dis,\A}_\at=H_\A x_\A+k_\B T[x_\A \ln x_\A+ (1-x_\A) \ln
(1-x_\A)], \\\\
    \Delta G^{\dis,\V}_\at=H_\V x_\V+k_\B T[x_\V \ln x_\V - (1+x_\V) \ln
(1+x_\V)].
\end{array}
\end{equation}

In alloy with fixed atomic composition $x_\A$,
the equilibrium concentration of vacancies $x_\V^\eq$
is determined by minimizing the formation Gibbs free energy:
\begin{equation}\label{Eq:G7}
    \left. \frac{\partial \Delta G^\dis_\at}{\partial
    x_\V} \right|_{x_\A}=\frac{\partial \Delta
    G^{\dis,\V}_\at}{\partial x_\V}=0.
\end{equation}
The manipulation of Eqs. (\ref{Eq:G6})-(\ref{Eq:G7})
leads to:
\begin{equation}\label{Eq:G8}\label{Eq:G9}\begin{array}{c}
    x_\V^\eq=\left [\exp \left(\frac{H_\V}{k_\B T}\right)-1 \right]^{-1},\\\\
    \left. x_\V^\eq\right|_{k_\B T \ll E_\V}\simeq \exp \left(-\frac{H_\V}{k_\B T} \right ), \\\\
    \Delta G^{\dis,\V}_\at(x_\V^\eq)= k_\B T \ln \left[1-\exp \left(-\frac{H_\V}{k_\B T}\right)\right], \\\\
    \left.\Delta G^{\dis,\V}_\at(x_\V^\eq)\right|_{k_\B T \ll E_\V} \simeq -k_\B T \exp \left(-\frac{H_\V}{k_\B T} \right).
\end{array} \end{equation}
To conclude, the ``true'' vacancies formation enthalpy $H_\V$ determines
the equilibrium concentration of vacancies with an Arrhenius-type
equation (see also Ref. \onlinecite{Korzh00}).
In the next section, we show that $H_\A$ resolves the solubility
in the case of a phase-separating alloy having no intermediate ground states.

\section*{Solubility}

\begin{figure}  \vspace{-5mm}
  \includegraphics[width=0.5\textwidth]{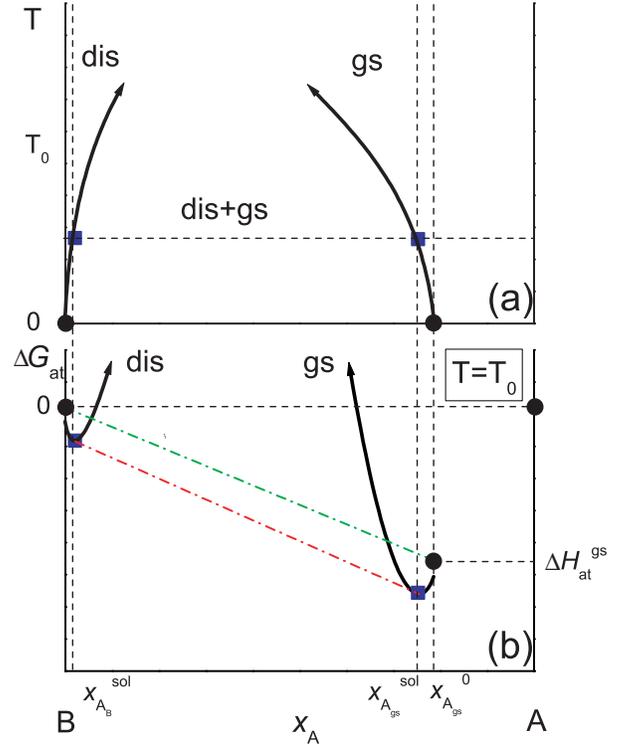}
  \vspace{-5mm}
  \caption{\small (Color online)
  (a) Temperature-concentration and (b) formation Gibbs free energy-concentration (at a given temperature $T_0$)
  graphs illustrating our solubility concepts.}\label{Fig:F-x}
\end{figure}

At a given temperature, the solubility of A-atoms in a B-solid
phase, $x_{\A_\B}^\sol$, is defined as the maximum homogenously
achievable concentration of A without the formation of a new phase
(Fig. \ref{Fig:F-x}(a)). The accurate calculation of
$x_{\A_\B}^\sol$ requires considering of the nearby ground state
(labeled as {\it``gs''}) with respect to the increase of $x_\sA$,
see the ``red line'' in Fig. \ref{Fig:F-x}(b). It is implemented by
minimizing the formation Gibbs free energy $\Delta G_\at^\mix(x)$ of
a mixture of (1) a disordered dilute solution of A-atoms and
vacancies in a B-rich solid phase at composition $x_{\A_\B}$ (the
``dis''-phase of the previous section), and (2) an on- or
off-stoichiometric ground state ``gs''-phase at composition
$x_{\A_\gs}$. The lever rule gives the fractions of the two phases:
\vspace{-4mm}
\begin{equation}\label{Eq:ComTang0}
  \begin{array}{c}
    \Delta G_\at^\mix(x)=\\\\
    \frac{x_{\A_\gs}-x}{x_{\A_\gs}-x_{\A_\B}} \Delta G^\dis_\at(x_{\A_\B})+
    \frac{x-x_{\A_\B}}{x_{\A_\gs}-x_{\A_\B}} \Delta G^\gs_\at(x_{\A_\gs}),
\end{array} \end{equation}
and the minimization is performed with respect to $x_{\A_\B}$ and
$x_{\A_\gs}$ ($x$ is the overall composition of A in the two-phase mixture, $x_{\A_\B}<x<x_{\A_\gs}$):
\begin{equation}\label{Eq:ComTang1}
    \frac{\partial \Delta G_\at^\mix}{\partial
    x_{\A_\B}}=0,\,\,\,\,\,\,\frac{\partial \Delta
    G_\at^\mix}{\partial
    x_{\A_\gs}}=0.
\end{equation}
Combining Eqs. (\ref{Eq:ComTang0}) and (\ref{Eq:ComTang1}) leads to
the usual {\it common-tangent} rule:
\begin{equation} \label{Eg:ComTang2}\begin{array}{c}
    \frac{\partial \Delta G^\dis_\at(x_{\A_\B})}{\partial
    x_{\A_\B}}= \frac{\partial
    \Delta G_\at^\gs(x_{\A_\gs})}{\partial
    x_{\A_\gs}}=\\\\
    \frac{\Delta G^\dis_\at(x_{\A_\B})-\Delta G_\at^\gs(x_{\A_\gs})}
    {x_{\A_\B}-x_{\A_\gs}}.
\end{array} \end{equation}
Substituting Eqs. (\ref{Eq:G5}) into Eqs. (\ref{Eg:ComTang2}), we obtain
\begin{equation}\label{Eq:Xsol1}
    x_{\A_\B}^\sol = \left [\exp
    \left(H_\sol/k_\B T\right)+1 \right ]^{-1},
\end{equation}
which approximates as an Arrhenius-type relation at low temperature:
\begin{equation}\label{Eq:Xsol2}
    \left. x_{\A_\B}^\sol\right|_{k_\B T
    \ll H_\sol } \simeq \exp \left(
    -H_\sol/k_\B T \right).
\end{equation}
The quantity $H_\sol$ is defined as
\begin{equation}\label{Eq:Esol2}
    H_\sol \equiv
    H_\A-\frac{\Delta G^\dis_\at(x^\sol_{\A_\B})-\Delta G_\at^\gs(x^\sol_{\A_\gs})}
    {x^\sol_{\A_\B}-x^\sol_{\A_\gs}}.
\end{equation}
The non-linear problem described by Eqs.
(\ref{Eq:Xsol1}-\ref{Eq:Xsol2}) and (\ref{Eq:Esol2}) can be
linearized in the low-solubility limit (labeled as {\it ``ls''}):
\begin{equation}\label{Eq:LS}
    ls: \left\{\begin{array}{c}
    \,x^\sol_{\A_\B}\simeq0,\,
    \Delta G^\dis_\at(x^\sol_{\A_\B})\simeq
    \Delta G^{\dis,\V}_\at(x_\V^\eq),\\\\
    x^\sol_{\A_\gs}\simeq
    x_{\A_\gs}^0,\,\,
    \Delta G_\at^\gs(x^\sol_{\A_\gs})\simeq
    \Delta H_\at^\gs, \end{array} \right.
\end{equation}
where $\Delta H_\at^\gs$ is the formation enthalpy
of the ground state ``gs''.
In the low-solubility limit, $H_\sol$  becomes:
\begin{equation} \label{Eq:EsolLS}
    H_\sol^\ls =
    H_\sol^{\ls,\nv}+\Delta
    G^{\dis,\V}_\at(x_\V^\eq)/x_{\A_\gs}^0,
\end{equation}
where $H_\sol^{\ls,\nv}$ is the non-vacancy contribution (labeled as {\it ``nv''}):
\vspace{-3mm}
\begin{equation} \label{Eq:EsolLSNV}
    H_\sol^{\ls,\nv} =
    H_\A-\Delta H_\at^\gs/x_{\A_\gs}^0.
\end{equation}
From the equilibrium vacancy concentration, Eq. (\ref{Eq:G9}),
the exponential part of Eqs. (\ref{Eq:Xsol1}-\ref{Eq:Xsol2}) becomes:
\begin{equation} \label{Eq:expV} \begin{array}{c}
    \exp \left(-H_\sol^\ls/k_\B T \right)= \exp \left(-H_\sol^{\ls,\nv}/k_\B T \right)\times \\\\
    \left[1-\exp \left(-\frac{H_\V}{k_\B T} \right) \right]^{-1/x_{\A_\gs}^0},
\end{array} \end{equation}
where the two contributions, non-vacancy and vacancy, are factorized.
The last expression indicates that the presence of vacancies effectively increases the solubility
by decreasing the number of host B-atoms in the solution.

\section{Interpretation of $H_\sol^{\ls,\nv}$}

\begin{figure}
  \vspace{-5mm}
  \includegraphics[width=0.45\textwidth]{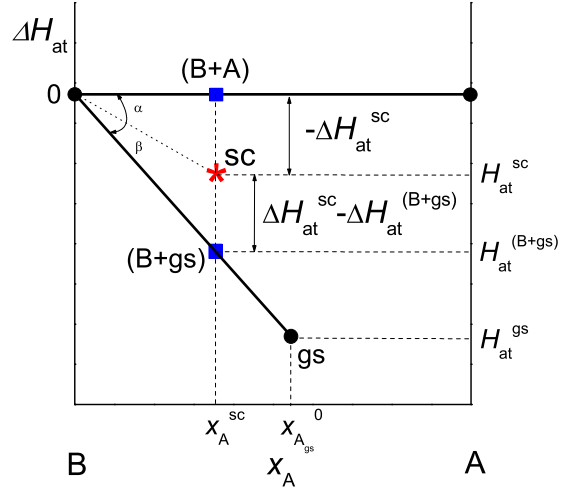}
  \vspace{-5mm}
  \caption{\small
  (Color online) The enthalpy-concentration graph demonstrating our concept of low-solubility formation enthalpy.
  A-pure, B-pure and "gs" are the ground states forming a convex hull.
  "sc" is a compound (supercell) with some intermediate enthalpy and composition.
  Compounds (B+A) and (B+gs) correspond to appropriate phase
  mixtures of the same general composition $x_\A^\sc$.} \label{Fig:E-x}
\end{figure}

For low-solubility calculations of non-interacting defects,
the framework can be implemented through first-principles calculation of $H_\sol^\ls$ (Eq. (\ref{Eq:EsolLS})).
It requires the knowledge of the enthalpy $\Delta H_\at^\gs$ and composition $x_{\A_\gs}^0$ of the ground-state ``gs'',
as well as the ``true'' defect formation enthalpies $H_\A$ and $H_\V$ (Eq. (\ref{Eq:H8})).

To capture the physical meaning of $H_\sol^{\ls,\nv}$,
let us consider an arbitrary
dilute solution ``sc'' at composition $x_\A^\sc$
(see Fig. \ref{Fig:E-x}).
The label ``sc'' indicates that the solution is generated as a {\it supercell} of the B-solid
upon insertion of defects, randomly distributed but not too close to avoid interactions.
The first part of Eq. (\ref{Eq:EsolLSNV}) can be rewritten according to Eq. (\ref{Eq:H9}) as:
\begin{equation}\label{Eq:Dis2}
    H_\A=\frac{\partial \Delta
    H_\at^\sc(x_\A^\sc)}{\partial
    x_\A^\sc}.
\end{equation}
The second part of Eq. (\ref{Eq:EsolLSNV}) becomes:
\begin{equation}\label{Eq:Dis1}
    \frac{\Delta
    H_\at^\gs}{x_{\A_\gs}^0}=
    \frac{\Delta
    H_\at^{(\B+\gs)}(x_\A^\sc)}{x_\A^\sc}=
    \frac{\partial \Delta
    H_\at^{(\B+\gs)}(x_\A^\sc)}{\partial x_\A^\sc},
\end{equation}
where $\Delta H_\at^{(\B+\gs)}(x_\A^\sc)$ is the
formation enthalpy of the mixture of the pure B-solid with
the ground state ``gs'' with overall composition $x_\A^\sc$
(point (B+``gs'') in Fig. \ref{Fig:E-x}).
Thus, we obtain:
\begin{equation}\label{Eq:Dis3}
    H_\sol^{\ls,\nv}=
    \frac{\partial \left[\Delta
    H_\at^\sc(x_\A^\sc)-
    \Delta
    H_\at^{(\B+\gs)}(x_\A^\sc) \right]}{\partial
    x_\A^\sc}.
\end{equation}
The comparison of Eq. (\ref{Eq:Dis2}) with Eq. (\ref{Eq:Dis3}) leads to the conclusion
that both $H_\A$ and $H_\sol^{\ls,\nv}$ are derivatives of supercell
formation energies with respect to A-composition.
For $H_\A$, the supercell formation enthalpy is determined with respect to pure A and B phases
(the distance between points ``sc'' and (B+A) in Fig. \ref{Fig:E-x}).
For $H_\sol^{\ls,\nv}$, the supercell formation enthalpy is determined
 with respect to B-pure and the ground state ``gs''
(the distance between points ``sc'' and (B+gs) in Fig.
\ref{Fig:E-x}). To conclude, $H_\A$ and $H_\sol^{\ls,\nv}$ are
characterized by the angles $\alpha$ and $\beta$ between the
B-``sc''/B-A and B-``sc''/B-``gs'' lines, respectively. In analogy
with the $H_\A$ definition in Eq. (\ref{Eq:H8}), $H_\sol^{\ls,\nv}$
can be considered as the ``low-solubility formation
enthalpy''.

The quantities $H_\A$ and $H_\sol^{\ls,\nv}$ are identical only for
phase-separating alloys having no intermediate ground states (i.e. ``gs''$\equiv$ A).
In this case the low-solubility can be formally determined by minimizing $\Delta G^{\dis,\A}_\at$
(Eq. (\ref{Eq:G6})) with respect to $x_{\A}$ (e.g. Ref. \onlinecite{Matysina}).
However, in the general case, the existence of ordered ground-states must be verified
so the appropriate formalism is used.
Generally, in ordering alloys $H_\A$ and $H_\sol^{\ls,\nv}$ differ,
and might even have different signs.

Note that if the ``sc'' point is below (B+gs) in Fig. \ref{Fig:E-x}
%($\Delta H_\at^\sc(x_\A^\sc)<\Delta H_\at^{(\B+\gs)}(x_\A^\sc)$ and
($H_\sol^{\ls,\nv}<0$)
the solubility expressions (\ref{Eq:Xsol1}-\ref{Eq:Xsol2}) are not valid,
and there must exist an undetected ground state (it might be ``sc'' itself)
with concentration lower than $x_{\A_\gs}^0$.
In this case, such undetected ground state should be used for the calculation of solubility
rather than initial ``gs'' \cite{RomanMgBA}.

The expression for non-binary low-solubility within the regular
solution model derived in Refs. [\onlinecite{SigliPROC,SanchezPRB}]
coincides with our derivation in the case of binary alloys without
vacancies and high-temperature contributions. This is because the
regular solution model corresponds to our model for the free energy
in case of dilute solution.

\section{Ground states of selected Titanium alloys}

As an example of our formalism, we calculate the solubility of a set
of metals in titanium. First, we explore the possible ground states
of the A-Ti systems (A=Ag,Au,Cd,Co,Cr,Ir,W,Zn) and then we apply the
construction described in the previous section.

The low temperature stability of A-Ti is performed by using our
high-throughput quantum calculations framework
\cite{Curtarolo2005Calphad,Curtarolo2003PRL,artSC21,artSC37}, based
on first-principles energies obtained with the {\small VASP}
software \cite{kresse1993}. We use projector augmented waves (PAW)
pseudopotentials \cite{bloechl994} and exchange-correlation
functionals as parameterized by Perdew-Burke-Ernzerhof \cite{PBE}
for the generalized gradient approximation (GGA). Simulations are
carried out with spin polarization except Ti-Co, at zero temperature
and pressure \cite{enthalpy-energy}, and without zero-point motion.
All structures are fully relaxed (shape and volume of the cell and
internal positions of the atoms). The effect of lattice vibrations
is omitted. Numerical convergence to within about 1 meV/atom is
ensured by enforcing a high energy cut-off (357 eV) and dense 6,000
{\bf k}-point meshes.

The number of crystal structures considered for the
calculations of each A-Ti system \textbf{is} 194. In addition to the
176 described in Ref. \onlinecite{Curtarolo2005Calphad}, we
included the following prototypes A5, A6, A7, A11,
Ca$_7$Ge, NbNi$_8$ (Pt$_8$Ti), V$_4$Zn$_5$, C36 and the whole
complete set of hcp-superstructures with up to 4 atoms/cell. The
whole process is performed in an automatic fashion through the
software {\small AFLOW} which generates the prototypes, optimize the
parameters, perform the calculations, correct possible errors, and
calculate the phase diagrams \cite{Curtarolo2005Calphad,AFLOW}.

\begin{figure}[htb]
%  \vspace{-5mm}
  \includegraphics[width=0.4\textwidth]{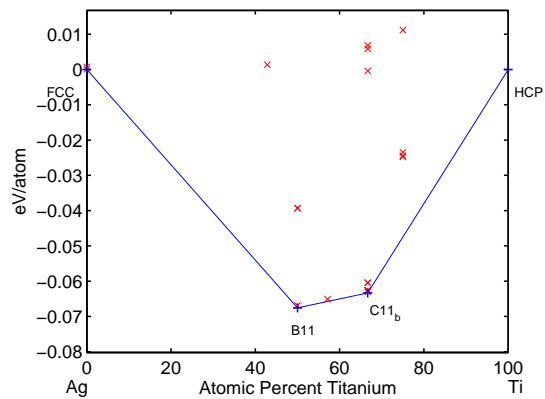}
  \vspace{-4mm}
  \caption{\small
  (Color online) AgTi (silver-titanium) ground state convex hull.}
  \label{FIG:AgTi_GS}
\end{figure}

\begin{figure}[htb]
  \vspace{-3mm}
  \includegraphics[width=0.4\textwidth]{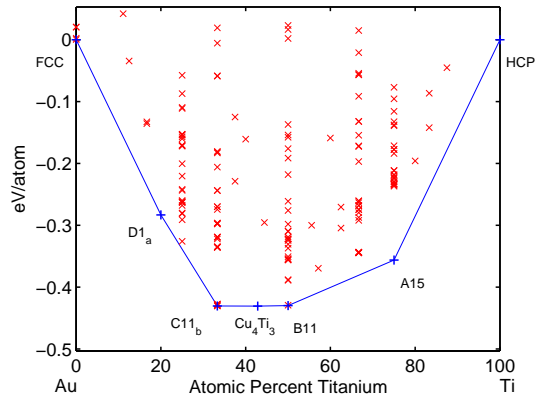}
  \vspace{-4mm}
  \caption{\small
  (Color online) AuTi (gold-titanium) ground state convex hull.}
  \label{FIG:AuTi_GS}
\end{figure}

\begin{figure}[htb]
  \vspace{-3mm}
  \includegraphics[width=0.4\textwidth]{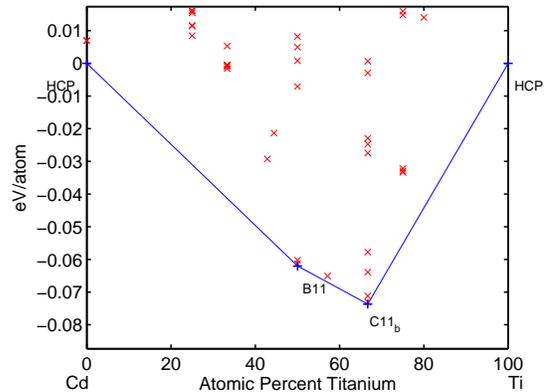}
  \vspace{-4mm}
  \caption{\small
  (Color online) CdTi (cadmium-titanium) ground state convex hull.}
  \label{FIG:CdTi_GS}
\end{figure}

\begin{figure}[htb]
  \vspace{-3mm}
  \includegraphics[width=0.4\textwidth]{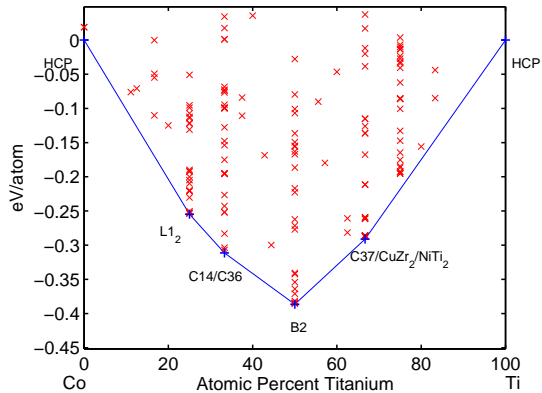}
  \vspace{-4mm}
  \caption{\small
  (Color online) CoTi (cobalt-titanium) ground state convex hull.}
  \label{FIG:CoTi_GS}
\end{figure}

\begin{figure}[htb]
  \vspace{-3mm}
  \includegraphics[width=0.4\textwidth]{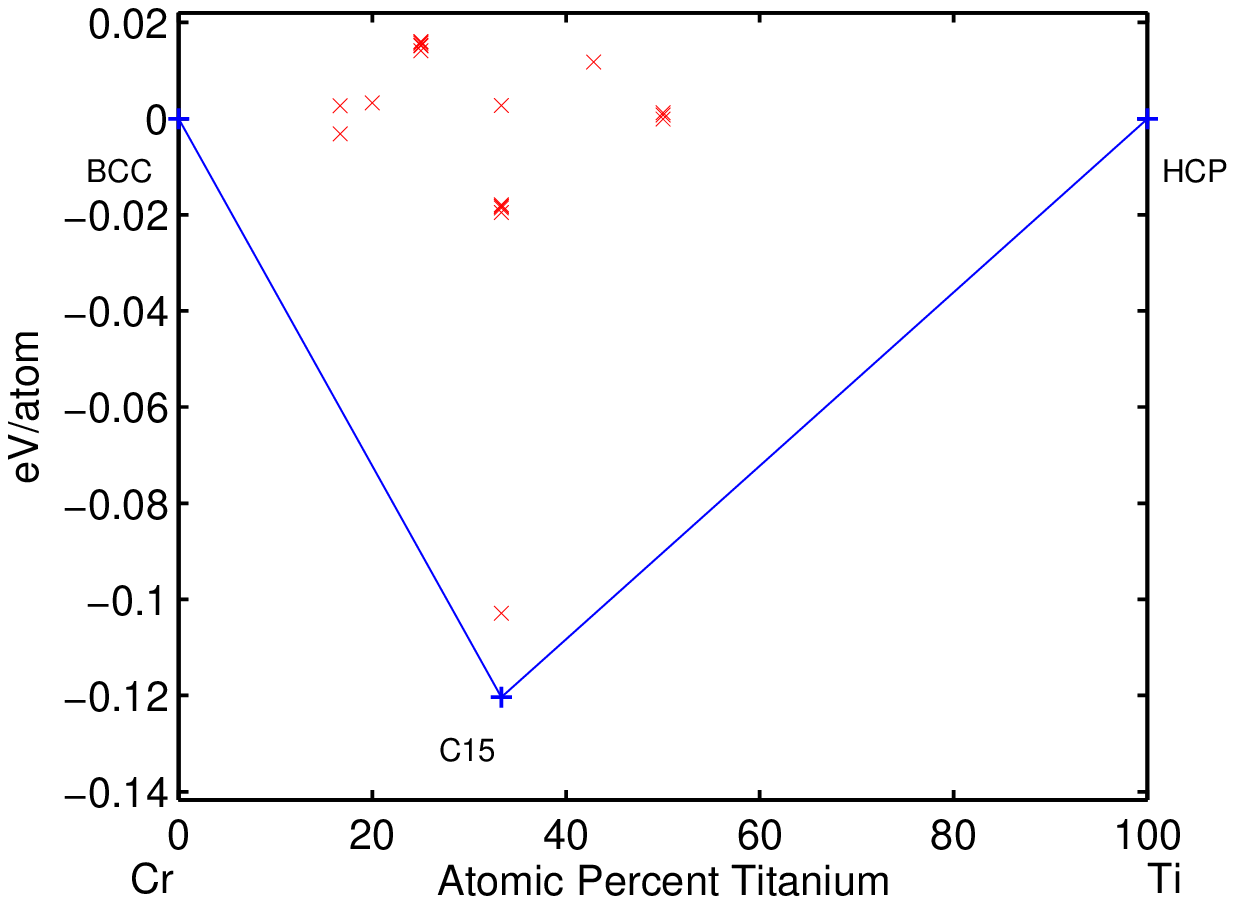}
  \vspace{-4mm}
  \caption{\small
  (Color online) CrTi (cromium-titanium) ground state convex hull.}
  \label{FIG:CrTi_GS}
\end{figure}

\begin{figure}[htb]
  \vspace{-3mm}
  \includegraphics[width=0.4\textwidth]{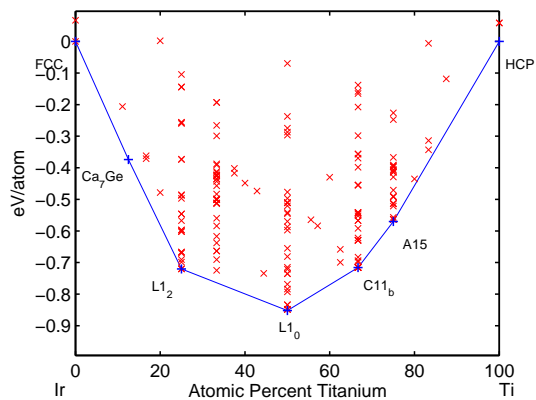}
  \vspace{-4mm}
  \caption{\small
  (Color online) IrTi (iridium-titanium) ground state convex hull.}
  \label{FIG:IrTi_GS}
\end{figure}

\begin{figure}[htb]
  \vspace{-3mm}
  \includegraphics[width=0.4\textwidth]{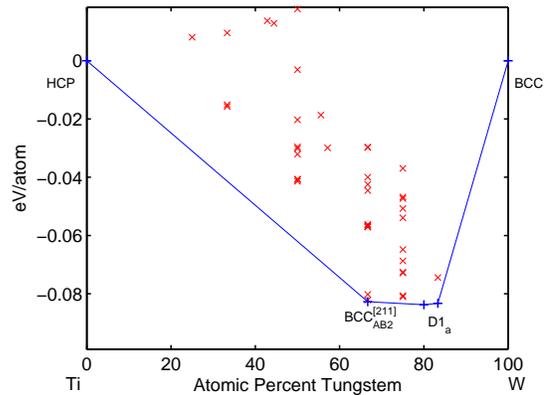}
  \vspace{-4mm}
  \caption{\small
  (Color online) TiW (titanium-tungsten) ground state convex hull.}
  \label{FIG:TiW_GS}
\end{figure}

\begin{figure}[htb]
  \vspace{-3mm}
  \includegraphics[width=0.4\textwidth]{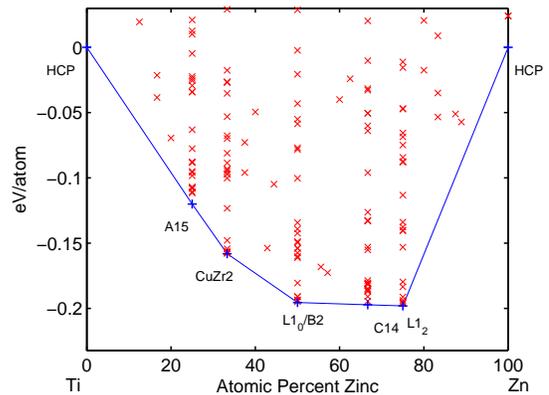}
  \vspace{-4mm}
  \caption{\small
  (Color online) TiZn (titanium-zinc) ground state convex hull.}
  \label{FIG:TiZn_GS}
\end{figure}

Our results of ground state calculation are presented in Figs.
\ref{FIG:AgTi_GS}-\ref{FIG:TiZn_GS} and Table \ref{Tab:GS}. The
correspondence {\it ab initio} versus experimental results is very
good and typical for this type of calculations
\cite{Curtarolo2005Calphad}.

We propose degenerate results for:
1) Co$_2$Ti, experimentally reported as C15,
but with {\it ab initio} formation energies of
-311.4 meV/at. and -304.0 meV/at. for
C14 and C15, respectively;
2) CoTi$_2$, experimentally reported as NiTi$_2$,
but with {\it ab initio} formation energies of
-291.2 meV/at., -287.3 meV/at., and -285.7 meV/at. for
C37, CuZr$_2$, and NiTi$_2$, respectively;
3) TiZn, experimentally reported as B2,
but with {\it ab initio} formation energies of
-195.4 meV/at. and  193.4 meV/at. for
L1$_0$ and B2, respectively.

We propose the novel results for: 1) Ir$_7$Ti,
experimentally reported as a two phases region above 500\DEG, but
with a possible {\it ab inito} low temperature ground state Ca$_7$Ge
with formation energy of -373.8 meV/at. 2) Ir$_2$Ti, experimentally
reported as a two phases region above 500\DEG, but with a possible
{\it ab initio} low temperature ground state C11$_b$, with formation
energy of -716.0 meV/at. 3) TiW$_2$, experimentally reported as a
two phases region above 500\DEG, but with a possible {\it ab initio}
low temperature ground state BCC$_{AB2}^{[211]}$, with formation
energy of -82.7 meV/at. (see notation for the prototype in Ref.
\cite{Curtarolo2005Calphad}). 4) TiW$_4$, experimentally reported as
a two phases region above 500\DEG, but with a possible {\it ab
initio} low temperature ground state D1$_a$, with formation energy
of -83.8 meV/at. 5) Ti$_3$Zn, experimentally not explored, but with
a possible {\it ab initio} low temperature ground state A15, with
formation energy of -120.0 meV/at.

In particular, the results indicate that in the Ir-Ti system the low
temperature Ir-rich part of the known phase diagram is not complete,
and that the Ti-W alloy has an ordering tendency at low
temperature, in contrast to common belief \cite{Pauling,BB}. In
addition, in Ti-Zn there must exist a Ti-rich compound with Ti
composition higher than the reported Ti$_2$Zn, \cite{Pauling,BB}.

\begin{table}[htb]
  \caption{\small
    Low temperature phases comparison chart for Ag-Ti, Au-Ti, Cd-Ti, Co-Ti, Cr-Ti, Ir-Ti, Ti-V, and Ti-Zn.
    Experimental data correspond to the lowest available temperature \cite{Curtarolo2005Calphad}.
    The ``$^*$'' indicates the energy values used in solubility calculations.}
  \label{Tab:GS} \small
  \begin{tabular}{c|c|c|c|c} \hline\hline
    Experimental             &  {\it Ab initio}    & $\Delta E^{\gs}_\at$  & Space   & Pearson \\
    (Refs. \onlinecite{BB,Pauling})  & result      & (meV/at.)           &  group\cite{IntTabs}    & \\ \hline
    \multicolumn{5}{c}{{\bf Ag-Ti}} \\\hline
    AgTi-B11                 & B11                 & -67.6           &  P4/nmm  & tP4     \\ \hline
    AgTi$_2$-C11$_b$         & C11$_b$             &  -63.3$^*$ &  I4/mmm  & tI6     \\ \hline
    \multicolumn{5}{c}{{\bf Au-Ti}} \\\hline
    Au$_4$Ti-D1$_a$          & D1$_a$              & -283.2          &  I4/m    & tI10    \\ \hline
    Au$_2$Ti-C11$_b$         & C11$_b$             & -430.4          &  I4/mmm  & tI6     \\ \hline
    two-phase region         & Au$_4$Ti$_3$-       & -430.6          &  I4/mmm  & tI14    \\
    above 500\DEG            & Cu$_4$Ti$_3$/tie    &                   &          &         \\ \hline
    AuTi-B11                 & B11                 & -429.8          &  P4/nmm  & tP4     \\ \hline
    AuTi$_3$-A15             & A15                 &  -356.1$^*$&  Pm$\bar{3}$n &  cP8  \\ \hline
    \multicolumn{5}{c}{{\bf Cd-Ti}} \\\hline
    CdTi-B11                 & B11                 & -62.0           &  P4/nmm  & tP4     \\ \hline
    CdTi$_2$-C11$_b$         & C11$_b$             &  -73.7$^*$      &  I4/mmm  & tI6     \\ \hline
    \multicolumn{5}{c}{{\bf Cr-Ti}} \\\hline
    Cr$_2$Ti-C15             & C15                 &  -120.3$^*$&  Fd$\bar{3}$m   & cF24    \\ \hline
    \multicolumn{5}{c}{{\bf Co-Ti}} \\\hline
    Co$_3$Ti-L1$_2$          & L1$_2$              & -254.8            & Pm$\bar{3}$m& cP4     \\ \hline
    Co$_2$Ti-C15             & C14                 & -311.4            & P6$_3$/mm & hP12   \\
                             & C15                 & -304.0            &  Fd$\bar{3}$m   & cF24    \\ \hline
    CoTi-B2                  & B2                  & -386.4            &  Pm$\bar{3}$m   & cP2     \\ \hline
    CoTi$_2$-NiTi$_2$        & C37                 &  -291.2$^*$&  Pnma    & oP12    \\
                             & CuZr$_2$            & -287.3            &  I4/mmm  & tI6     \\
                             & NiTi$_2$            & -285.7            &  Fd$\bar{3}$m   & cF96    \\ \hline
    \multicolumn{5}{c}{{\bf Ir-Ti}} \\\hline
    two-phase region         & Ir$_7$Ti-           & -373.8            &  Fm$\bar{3}$m   & cF32    \\
    above 500\DEG            & Ca$_7$Ge            &                   &          &         \\ \hline
    Ir$_3$Ti-L1$_2$          & L1$_2$              & -720.3            & Pm$\bar{3}$m& cP4     \\ \hline
    IrTi-$\delta$            & L1$_0$              & -851.8            & P4/mmm   & tP4     \\ \hline
    two-phase region         & Ir$_2$Ti-           & -716.0            &  I4/mmm  & tI6     \\
    above 500\DEG            & C11$_b$             &                   &          &         \\ \hline
    IrTi$_3$-A15             & A15/tie             &  -570.3$^*$& Pm$\bar{3}$n& cP8     \\ \hline
    \multicolumn{5}{c}{{\bf Ti-W}} \\\hline
    two-phase region         & TiW$_2$-            &  -82.7$^*$ & P$\bar{3}$m1    & hP3     \\
    above 500\DEG            & BCC$_{AB2}^{[211]}$ &                   &          &         \\ \hline
    two-phase region         & TiW$_4$-            & -83.8             &  I4/m    & tI10    \\
    above 500\DEG            & D1$_a$              &                   &          &         \\ \hline
    \multicolumn{5}{c}{{\bf Ti-Zn}} \\\hline
    non explored             & Ti$_3$Zn-A15        &  -120.0$^*$& Pm$\bar{3}$n& cP8     \\ \hline
    Ti$_2$Zn-CuZr$_2$        & CuZr$_2$            & -158.0            &  I4/mmm  & tI6     \\ \hline
    TiZn-B2                  & L1$_0$              & -195.4            & P4/mmm   & tP4     \\
                             & B2                  & -193.4            &  Pm$\bar{3}$m   & cP2     \\ \hline
    TiZn$_2$-C14             & C14/tie                 & -197.2            &  P6$_3$/mm & hP12  \\ \hline
    TiZn$_3$-L1$_2$          & L1$_2$              & -198.0            &  Pm$\bar{3}$m & cP4   \\ \hline
%    TiZn$_5$-unknown         & unavail.            & -                 &          &         \\ \hline
%    Ti$_3$Zn$_{22}$          & unavail.            & -                 &          &         \\ \hline
%    TiZn$_{16}$              & unavail.            & -                 &          &         \\
   \hline\hline \end{tabular}
  \vspace{-0.5mm}
\end{table}

\section{Results and discussion of solubility in Ti alloys}
\vspace{-2mm}

\begin{figure}
  \vspace{-5mm}
  \includegraphics[width=0.5\textwidth]{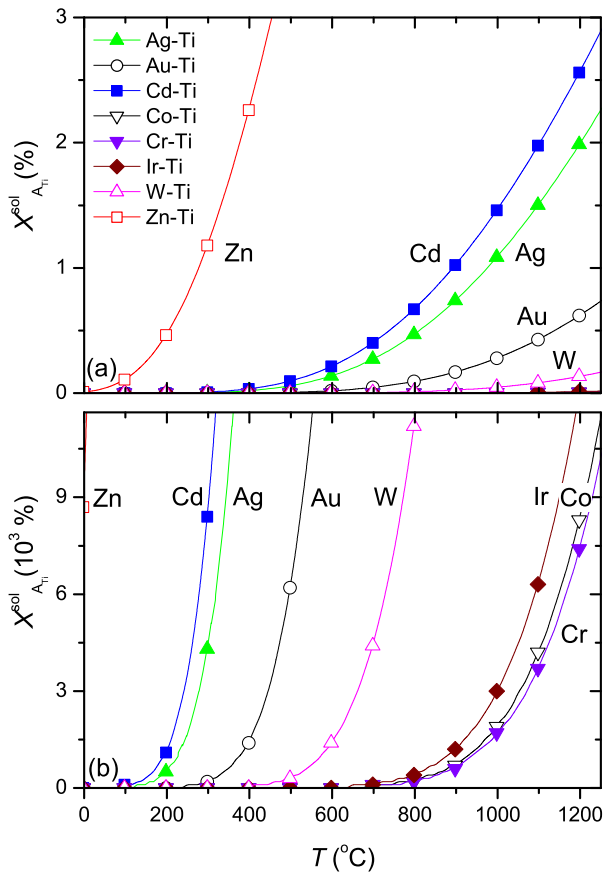}
  \vspace{-12mm}
  \caption{\small
    (Color online)
    The solubilities of eight considered transition metals
    in titanium as functions of temperature (calculated through Eqs. (\ref{Eq:Xsol1},\ref{Eq:EsolLS})).
    Panel (b) is a magnified version of panel (a), to visualize the values for Ir, Co, and Cr.}
  \label{FIG:Sol_T_theor}
\end{figure}

\begin{figure}
  \vspace{-5mm}
  \includegraphics[width=0.5\textwidth]{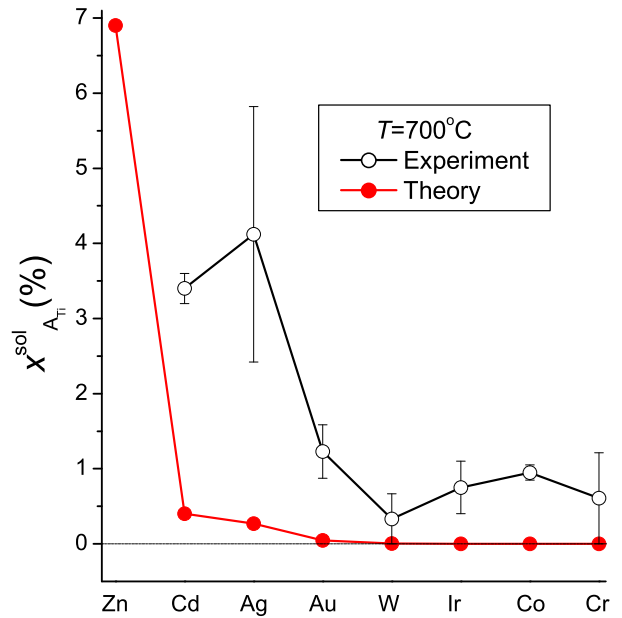}
  \vspace{-5mm}
  \caption{\small
  (Color online) Comparison of experimental\cite{Pauling,BB}
    and theoretical solubilities of eight considered transition metals
    in titanium at T=700\DEG. From left to right, the elements are
    ordered from low to high low-solubility formation enthalpy (and
    correspondingly theoretical solubility) The error bars characterize
    the scattering of data measured in different experiments. We could
    not find the experimental data for solubilities in Ti-Zn.}
  \label{FIG:Sol_theor_exper}
\end{figure}

The ``raw'' formation enthalpies of Ti alloys with one
substitutional atom or a vacancy were obtained through Eqs.
(\ref{Eq:H3}) considering $3\times3\times3$ supercells of the hcp
Ti. Supercell dimensions were chosen to limit the defect-defect
interactions, being their distance at least three times larger than
the nearest neighbor Ti-Ti bond. The results are presented in Table
\ref{Tab:FormEn}. Solubilities temperature dependencies are
presented in Fig. \ref{FIG:Sol_T_theor}, while Fig.
\ref{FIG:Sol_theor_exper} illustrates a comparison of experimental
and theoretical data at T=700\DEG.

\vspace{-3mm}
\begin{table}[thb]
  \caption{\small
    ``Raw'' formation enthalpies $H_{\A_\Ti}$,
    $H_{\V_\Ti}$ (from Eqs. (\ref{Eq:H1}-\ref{Eq:H3}) at
    zero pressure), ``true'' formation enthalpies $H_\A$,
    $H_\V$ (from Eq. (\ref{Eq:H8}) at zero pressure) and
    low-solubility (non-vacancy) formation enthalpy
    $H_\sol^{\ls,\nv}$ (from Eq. (\ref{Eq:EsolLSNV}) at
    zero pressure) of substitutional A-defects (A=Ag,Au,Cd,Co,Cr) and
    vacancies (V). The elements are ordered from low to high
    {\it ``low-solubility formation enthalpy''}. 
    All quantities are in eV units.}\label{Tab:FormEn}
  \vspace{0.5mm}
  \begin{tabular}{c|ccc} \hline\hline
    A   &   $H_{\A_\Ti}$ &   $H_\A$  &
    $H_\sol^{\ls,\nv}$ \\ \hline
    Zn  &   6.405   &   -0.262  &   0.218   \\
    Cd  &   7.267   &   0.24    &   0.461   \\
    Ag  &   5.412   &   0.304   &   0.494   \\
    Au  &   3.882   &   -0.781  &   0.643   \\
    W   &   -4.303  &   0.714   &   0.838   \\
    Ir  &   -2.059  &   -1.14   &   1.141   \\
    Co  &   1.138   &   0.316   &   1.19    \\
    Cr  &   -0.674  &   1.024   &   1.205   \\\hline
    \multicolumn{2}{c}{$H_{\V_\Ti}=10.002$,} &
    \multicolumn{2}{c}{$H_\V=2.068$} \\ \hline\hline
\end{tabular} \end{table}

Table \ref{Tab:FormEn} shows negative ``true'' formation enthalpies
$H_\A$ for A=Zn, Au, and Ir. This indicates that calculation of Zn,
Au, and Ir solubilities in Ti is not faceable without considering
nearby intermetallic ground states.

Figures \ref{FIG:Sol_T_theor}-\ref{FIG:Sol_theor_exper} show that
the highest theoretical solubilities in Ti occurs for Zn, Cd, Ag,
and Au as consequence of their low solubility formation enthalpies
(see Table \ref{Tab:FormEn}). This high solubility has also been
observed experimentally for Cd, Ag, and Au, whereas, to our best
knowledge, solubility of Zn does not seem to have been studied. High
solubility of late transition metals Zn, Cd, Ag, and Au in the early
transition metal titanium is due to the substantial localized
stability provided by the filling tendency of the $d$-band of Ti
(Ref. \onlinecite{Friedel}).

Because of the very high formation enthalpy of vacancies in Ti
(reported in Table \ref{Tab:FormEn}), the vacancy equilibrium
concentration was found to be very low at all considered
temperatures  (e.g. $x_\V^\texttt{eq}<10^{-6}$ at $T < 1300$\DEG ).
Thus, the effect of vacancies on the solubilities of Ti
alloys should be negligible.

Although, theoretical and experimental results follow similar
trends, theoretical solubilities are considerably smaller for most
of the considered alloys. A similar discrepancy was also observed
for Al-Zr in Ref. \onlinecite{SanchezPRB}. The discrepancy could be
due to shortcomings of theory and/or experiment. The main
approximations of our model consist of (a) neglecting the defect
interaction, (b) neglecting the spatial defect correlation and (c)
assuming low concentration of defects. However, as such
approximations are somehow related, subsequent solubility
calculations suggesting low values validate the assumptions (the
mean-field approximation neglects the interatomic positional
correlations but it should work well when the deviation from
complete stoichiometric (pure Ti-solid in our case) is small - and
so is the solubility in our case - see Sec. 19 in Ref.
\onlinecite{KrivSm64}.) For all considered alloys except Ti-Zn, we
found that at intermediate temperatures the calculated solubilities
are small enough for our approximations to be valid, although
smaller than experiment, as mentioned before. In
Ref.\onlinecite{SanchezPRB}, the authors added the defect
interactions through the Cluster Expansion method but without
increasing the solubility considerably. Actually, our formalism
includes the solute-solvent ordering tendency by considering the
real intermetallic ground state other then the pure Ti-solid. Thus,
we conclude that our approximations (or those of the cluster
expansion method) are not responsible for the theory-experiment
solubility discrepancy.

In case of Ti-Zn, the formation enthalpy is very low (see Table
\ref{Tab:FormEn}). Correspondingly, the theoretically solubility at
intermediate temperatures is very high, violating the assumptions of
the model. Thus, for Zn in Ti, a more precise solute interaction and
correlation parameterizations is required, which can be obtained by
using, for example, Cluster Expansion and Monte Carlo simulations.
Hence, the presented formalism should be considered as an
appropriate starting point to determine if more computationally
expensive formalisms are needed.

The other approximation of our model is the assumed independence of
our model energy and volume parameters on temperature. The
dependence can be caused by non configurational degrees of freedom,
like vibrational or anharmonicity (magnetic ordering is not actual
for considered alloys). However, theory-experiment solubility
discrepancy is observed even at low enough temperatures (e.g. $T
\leq \Theta _\D(\texttt{Ti})=$374-385 K \cite{Sundman01}), where the
vibrational contribution to the free energy is not important. In
fact, vibrational entropy is substantially smaller than the
configurational contribution \cite{AxelRMP}, so its inclusion can
not modify solubility results much. In fact, even in Ref.
\onlinecite{SanchezPRB}, the authors added phonon contribution
without increasing the solubility considerably.

On the other side, the experimental equilibrium solubility tends
usually to be overestimated. In fact, the formation of metastable
and/or unstable states which are subsequently frozen at low
temperatures, can make solubility measurements very challenging. In
such scenarios, the measured solubility may correspond to spinodal
concentration rather then actual binodal concentration or simply
characterize the frozen out of equilibrium solubility remaining from
the initial specimen preparation at higher temperature. Besides, the
segregation of defects into grain-boundaries, especially in
multicrystalline samples prepared through non optimal cooling
dramatically affect the amount of frozen defects and solutes.

\section{Conclusions}
Based on the statistical-thermodynamic theory of dilute lattice gas,
we developed an approach to calculate the atomic bulk solubility in
alloys. The advantage of our approach consists in considering all
the appropriate ground states rather than the pure species. It was
shown that the low-solubility follows simple Arrhenius-type
temperature dependence determined by a {\it ``low-solubility
formation enthalpy''}. This quantity is defined as the composition
derivative of the compound formation enthalpy with respect to nearby
ground states. ``Low-solubility formation enthalpy'' coincides with
the usual defect formation enthalpy only in the case of a
phase-separating alloy having no intermediate ground states and
vacancies. The key quantities of our model can be directly obtained
by first-principles calculations of by fitting experimental
temperature solubility dependence. Generalization of our model to
intermediate phases and/or to multicomponent, multisublattice,
interstitial-substitutional alloys is straightforward.

As examples, we applied the framework for a set ot eight Ti alloys
A-Ti (A=Ag,Au,Cd,Co,Cr,Ir,W,Zn). We have found that the highest
solubility for Zn, Cd, Ag, and Au is in qualitative
agreement with available experimental data and band structure
expectations. The quantitative differences between the theory and
experiment observed in the present and other similar studies are
discussed.

In conclusion, our formalism is correct in the limit of low-solubility and should be considered as an appropriate
starting point for determining if more computationally expensive formalisms are otherwise needed.

\section*{Acknowledgements}
We acknowledge Wahyu Setyawan, Mike Mehl, and Ohad Levy for fruitful discussions.
This research was supported by ONR (Grant No. N00014-07-1-0878)
and NSF (Grant No. DMR-0639822)
We thank the Teragrid Partnership (Texas Advanced Computing Center,
TACC) for computational support (MCA-07S005).


\begin{thebibliography}{*}
\bibitem{enthalpy-energy}
  So that by calling enthalpy we actually mean energy in this section.
\end{thebibliography}
\end{document}